\documentclass[pra,12pt]{revtex4-2}

\usepackage[a4paper, includefoot,
			left=1.5cm,
			right=1.5cm,
			top=2cm,
			bottom=2cm,			
			headsep=1cm,
			footskip=1cm]{geometry}
\usepackage{graphicx}

\newcommand{\podd}{$\cal P$-odd~}
\newcommand{\todd}{$\cal T$-odd~}

\begin{document}


\author{Timur Isaev}%
 \email{timur.isaev.cacn@gmail.com}
\affiliation{
Chemistry Dept., M. Lomonosov Moscow State University, Moscow 119991, Russia 
}%

\author{Dmitrii Makinskii}
\email{makinskii\_dm@pnpi.nrcki.ru}
\affiliation{%
 NRC ''Kurchatov Institute'' - PNPI, Orlova Roscha, 1, 188300 Gatchina, Russia  \\
 Chemistry Dept., M. Lomonosov Moscow State University, Moscow 119991, Russia 
}%

\author{Andrei Zaitsevskii}
\email{zaitsevskii\_av@pnpi.nrcki.ru}
\affiliation{%
 NRC ''Kurchatov Institute'' - PNPI, Orlova Roscha, 1, 188300 Gatchina, Russia  \\
 Chemistry Dept., M. Lomonosov Moscow State University, Moscow 119991, Russia 
}

\title{ Radium-containing molecular cations amenable for laser cooling}
\date{\today}
\pacs{31.30.-i, 37.10.Mn, 12.15.Mm, 21.10.Ky}
\begin{abstract}
Recently a new wave of interest to spectroscopy of radioactive compounds has raised due to 
successful applications of the new experimental ISOL/CRIS technique to optical spectroscopy of radium monofluoride
molecules. This opens great prospects to searches of the effects connected with ``new physics'' which point on 
deviations of the physical laws from those established by Standard Model of elementary particles. Considerable 
advantages in experiments for search for such effects would be provided by the abilities
to trap the working molecules, their
laser-coolability and presence of co-magnitometer molecular states and also a variety of accessible  
nuclear isotopes of the elements in the molecule with different 
nuclear multipole moments.
Laser-coolable polyatomic molecular ions containing Ra nuclei shall meet the above criteria and in the present article 
we are  considering a number of prospective molecular ions and calculating their electronic structure and molecular properties.
\end{abstract}
\maketitle

\section{Introduction}
Molecules containing radium nuclei offer great prospects in a number of research fields, including
search for ``new physics'' outside of the Standard Model of elementary particles. 
These prospects are related, on the one hand, to the extreme sensitivity of molecules containing heavy nuclei to the effects of 
time (\todd) and space (\podd) parity violation \cite{flambaum:1985, kozlov:1995,berger:2004a, acme:11, 
Kozyryev:2017b, Isaev:17,Gaul:2017} and
on the other hand to a considerable enhancement of nuclear \podd and \todd moments in octupole-deformed nuclei 
\cite{Auerbach:96, Spevak:97}. Radium isotopes possess a large variety of nuclear properties, including 
a wide range of nuclear spins from 0 ($^{226}$Ra) to 5/2 (e.g. in $^{221}$Ra), octupole 
deformation (in isotopes $^{223}$Ra and $^{225}$Ra) and half-live time from 1599 years for $^{226}$Ra 
to 4.7 seconds for $^{209}$Ra. Recently a new direction in molecular spectroscopy, namely spectroscopy
of short-lived radioactive molecules, has been established \cite{GarciaRuiz:2020} and the first vibrational 
spectra measurements have
been performed on radium monofluoride (RaF) molecule. As experimental information for this and other molecules 
containing radioactive nuclei was scarce, 
\emph{ab initio} electronic structure calculations \cite{Isaev:10a,Isaev:13,Zaitsevskii:2022}
 were  the main source of valuable information to design effective experimental scheme. 
 One of the important predictions following from the electronic structure calculations was
 good prospects for  laser-coolability of
 RaF, which paves the way for high-precision spectroscopic studies. Laser-coolability is also important in experiments
 with molecular ions \cite{Lien:2011}, where it allows to perform efficient compression of population 
 of vibrational and rotational degrees
 of freedom \cite{Lien:2014,Stollenwerk:2020}. Molecular ions also have an attractive feature, that they can be easily trapped 
 and then considerable trapping time can be used for optical manipulations. For example, in experiments on search of 
 permanent electric dipole moment (EDM) of an electron in HfF$^+$ molecular cation 
 \cite{Cairncross:2017}   the trapping time was ca.~1$s$, i.e.  sufficient to perform vibrational-rotational
 cooling even with optical cooling loop  involving highly-forbidden electronic transitions.
 In view of considerable progress of molecular spectroscopy
 of radioactive species and attractive features of radium-containing compounds for high-precision spectroscopy 
(see also \cite{Zakharova:2021, Zakharova:2022})
 we are considering a number of radium-containing molecular ions which 
  could possess highly-diagonal Franck-Condon matrices
 for transitions between ground and low-lying
 excited electronic states. Considered compounds can be produced using the technique similar to that used in \cite{Fan:2021}. We analyze electronic structure of suggested compounds from the point of
 laser-coolability 
(which requires quasi-diagonality of Franck-Condon matrix for cooling vibronic transitions \cite{DiRosa:2004}) and
is an essential prerequisite for ultra-precision spectroscopy.

\section{Calculation details and discussion}

All calculations reported below employ the relativistic electronic structure model similar to that used in Ref.~\cite{Isaev:21}. The model is defined by the accurate two-component shape-consistent pseudopotential
 of Ra (for discussion on the theory of RECP employed
see \cite{Mosyagin:10a,Mosyagin:16}) replacing 60 inner core electrons \cite{ourGRECP}. 
Light-element atoms are described in all-electron manner; the effects of relativity are accounted for through so-called empty-core pseudopotentials~\cite{Mosyagin:21,ourGRECP}. All RECPs
are optimized to describe valence (rather than subvalence and/or core) electronic shells and incorporate implicitly the effects of finite nuclear size and Breit electron-electron interactions.

Ground-state equilibrium geometries and energetics are determined using two-component non-collinear density functional theory \cite{Wullen:10} with the ``non-empirical'' PBE0 approximation for the exchange-correlation functional \cite{Adamo:99}. The components of one-electron (pseudo) spinors are expanded in the Gaussian basis comprising the primitive $(10s9p9d4f)$ set on the Ra atom and triple zeta quality sets \cite{Shaefer:94,Basissetexchange:19}) extended by double sets of polarization functions on other atoms; for first-row atoms, single sets of diffuse functions are added as well (see Supplementary materials). The total basis set sizes were thus $(12s7p2d)/[6s4p2d]$ for C, N, F and $(5s2p)/[3s2p] $ for H. The relativistic effects associated with light atoms were neglected at the DFT stage. The resulting geometries and Ra$^+$ binding energies are listed in Table \ref{dft}. It is worth noting that the obtained equilibrium parameters for RaNCH$^+$ agree well with those from relativistic CCSD(T) calculations (the Ra--N equilibrium  separation 2.905~\AA, dissociation energy  
$6.89\cdot10^{3}$~cm$^{-1}$,  see below).

\begin{table}[h!]\caption{Ground-state equilibrium geometries (bond lengths in \AA, angles in degrees) and dissociation energies of ionic complexes according to the two-component relativistic DFT / PBE0 calculations. 
}\label{dft}
\begin{center}
\begin{tabular}{lcccccc}
\hline \\ 
Complex          & Symmetry & $R_e$(Ra--X$^{a)}$)  & Attached ligand geometry & (deformation) &  $D_e$, 10$^3$ cm$^{-1}$ \\
\\
RaNH$_3^+$   & $C_{3v}  $   & 2.872         &$R$(N--H)=1.018  & ($+0.005$ )  &  8.13    \\
             &              &                &\angle{}RaNH=113.43$^\circ$& \\
             &              &                &\angle{}HNH=105.24$^\circ$& ($-1.43^\circ$) \\
\\
RaNCCH$_3^+$ & $C_{3v}  $   & 2.799         &$R$(N--C)=1.153  & ($-0.000$)  & 9.48 \\
             &              &                &$R$(C--C)=1.443  & ($-0.008$) \\
             &              &                &$R$(C--H)=1.091  & ($+0.001$) \\
             &              &                &\angle{}CCH=109.65$^\circ$&  ($-0.46^\circ$) \\
\\
RaNCH$^+$    &$C_{\infty v}$& 2.881          &$R$(N--C)=1.1470 \AA  & ($-0.003$) & 7.03 \\
             &              &                &$R$(C--H)=1.0726 \AA  & ($+0.005$) \\ 
\\
RaCNH$^+$    &$C_{\infty v}$& 3.073         &$R$(C--N)=1.157  & ($-0.012$ )$^{b)}$ & 6.68 $^{c)}$ \\
             &              &                &$R$(N--H)=1.002  & ($+0.006$ )$^{b)}$ & \\
\\
RaNCCN$^+$   &$C_{\infty v}$& 2.931         &$R$(C--N$_{\rm Ra}$)=1.154 & ($-0.001$)    & 4.23 \\
             &              &                &$R$(C--C)= 1.374   & ($-0.002$) \\
             &              &                &$R$(C--N$_{\rm term}$)=1.156& ($+0.001$)  & \\
\\
RaFH$^+$     & $C_{\infty v}$& 2.778 &$R$(F--H)=0.928& ($+0.009$) & 4.17 \\ \\
\hline \\ 
\end{tabular}
\end{center}
$^{a)}$ neighbor atom $^{b)}$ with respect to HNC, $^{b)}$ with respect to  Ra$^+$ + HNC 
\end{table} 

In all cases, the complex formation does not lead to significant deformation of the ligand
in comparison to its free form; for instance, the deformation of the free HCN molecule from its equilibrium geometry to that found for the ligand in the RaNCH$^+$ complex increases its energy by less than 40 cm$^{-1}$. The Ra$^+$ binding energies are rather moderate (the largest value, 9.48$\cdot 10^3$ cm$^{-1}$, is predicted for RaNCCH$_3^+$). The equilibrium structures of compounds have 
symmetry of the free ligand.

Excited state calculations for relatively stable complexes are performed within the standard Fock space relativistic coupled cluster (FS RCC) scheme  \cite{Visscher:01} using the closed-shell Hartree--Fock determinant of the doubly charged molecular cation as the Fermi vacuum state and restricting the cluster operator expansion to single and double excitations (FS RCCSD). For all systems under study, model spaces are constructed by placing the single electron on one of 18 lowest-energy molecular spinors corresponding to $7s_{1/2}$, $6d_{3/2,\,5/2}$, and $7p_{1/2,\,3/2}$ atomic spinors of radium ion at the dissociation limit. The employed basis for Ra is the $[10s\,9p\,9d\,7f\,4g\,3h]$ subset of the basis from Ref.~\cite{Isaev:21}; its important feature consists in using ANO-like high-angular-momentum functions optimised in scalar relativistic calculations. For hydrogen and first-row atoms we use correlation-consistent basis sets
cc-pVTZ and aug-cc-pVTZ \cite{Dunning:89,Kendall:92}, respectively. The $5s5p$ subshells of Ra as well as $1s$ shells of first-row atoms are frozen after the Hartree--Fock stage.

Crude estimates for squared vertical transition dipoles are obtained using the model-space parts of the wavefunctions as $|\langle \psi_i^{\perp\!\perp}\vert d_\eta \vert\psi_f\rangle|\times|\langle \psi_f^{\perp\!\perp}\vert d_\eta \vert\psi_i\rangle|$, where $\psi_i,\;\psi_f$ and  $\psi_i^{\perp\!\perp},\;\psi_f^{\perp\!\perp}$ are respectively right and left eigevectors of the 
non-Hermitian
FS RCC effective Hamiltonian corresponding to the initial ($i$) and final ($f$) states and $d_\eta$ stands for a component of the dipole moment operator. Due to the complete neglect of contributions to dipole matrix elements from the configurations outside of the model space, these estimates should be considered only as semiquantitative~\cite{Zaitsevskii:20}.

The FS RCC calculations are performed with the EXP-T package \cite{EXPT:20} which uses the solutions of spin-orbit-coupled Hartree-Fock equations and transformed molecular integrals exported by the DIRAC19 program system \cite{DIRAC:19,DIRAC:20}. The DIRAC19 code is also employed for relativistic single-reference CCSD(T) calculations.

\begin{table}[h!]\caption{FS RCC vertical excitation energies $T_{\rm vert}$ and squared transition dipoles  RaNCH$^+$ computed at the RDFT/PBE0 ground state geometries. 
\label{verti}}
\begin{center}
\begin{tabular}{lcccccc}
\hline \\ 
           & State & $T_{\rm vert}$, $10^{3}$ cm$^{-1}$  & $\sum |d^2|\;^{a)}$, a.u. \\
\\
RaNCH$^+$   &    ($X$)1/2         &      0      \\    
             &    (1)3/2          &   11.36     & 0.02 \\  
             &    (2)1/2          &   11.74     & 0.10 \\    
             &    (2)3/2          &   12.70     & 0.01 \\   
             &    (1)5/2          &   12.89     & 0 \\
             &    (3)1/2          &   16.10     & 3.45 \\    
             &    (4)1/2          &   20.08     & 7.73 \\    
             &    (3)3/2          &   22.38     & 8.55 \\    
             &    (5)1/2          &   29.82     & 2.07 \\    
              \\
RaNH$_3^+$   &    $(X)E_{1/2}$    &       0  \\
             &    $(1)E_{3/2}$    &   11.75     &  0.01   \\ 
             &    $(2)E_{1/2}$    &   13.05     &  
                                                             $<\!\!10^{-2}$\\
             &    $(3)E_{1/2}$    &   13.43     &  0.13   \\  
             &    $(2)E_{3/2}$    &   14.26     &  0.02  \\ 
             &    $(4)E_{1/2}$    &   17.07     &  4.16   \\ 
             &    $(5)E_{1/2}$    &   20.58     &  7.09  \\ 
             &    $(3)E_{3/2}$    &   22.62     &  8.79   \\ 
             &    $(6)E_{1/2}$    &   25.12     &  2.88   \\ 
\\
 RaNCCH$_3^+$ &    $(X)E_{1/2}$    &       0  \\
             &    $(1)E_{3/2}$    & 11.61      &  0.03   \\ 
             &    $(2)E_{1/2}$    & 12.26      &  0.23   \\ 
             &    $(3)E_{1/2}$    & 13.01      &  
             %
                                                             $<\!\!10^{-2}$\\
             &    $(2)E_{3/2}$    & 13.15      &  0.05   \\ 
             &    $(4)E_{1/2}$    & 16.20      &  3.99   \\ 
             &    $(5)E_{1/2}$    & 19.89      &  7.74   \\ 
             &    $(3)E_{3/2}$    & 21.93      &  8.73  \\ 
             &    $(6)E_{1/2}$    & 28.62      &  1.37   \\ 
\\  \hline
\end{tabular}
\end{center}
$^{a)}$ sum of squared dipoles for transitions from a single excited state component to both components of the ground state.
\end{table}

The results are summarized in Table~\ref{verti}. Four lowest vertical excitations of each molecular anion are localized approximately in the same energy range as $7s-6d$ excitations of the free atomic ion Ra$^+$ (12.08$\times10^3$ 
and 13.74$\times10^3$ wavenumbers) and are very weakly allowed or 
even effectively forbidden,  as it should be for quasi $s-d$ transitions,
due to rather weak bonding of Ra$^+$ with ligands (unlike 
the
cases of RaF \cite{GarciaRuiz:2020},
RaCl \cite{Isaev:21} and RaOH \cite{Isaev:17} where strong chemical bonding of Ra
leads to strong $s-p$/$p-d$ hybridisation of unpaired-electron spinors of corresponding low-lying
electronic states).
The fifth excited state, (3)1/2 or (4)$E_{1/2}$, also correlating with Ra$^+$ (6d) at the dissociation limit, still lies well below the area of  $7s-7p$ transitions of Ra$^+$
( 21.35$\times10^3$ and  26.21$\times10^3$ 
wavenumbers) but the transition dipole matrix element between this state and the ground one is large, indicating a significant admixture of $p-$ components.   
The lack of Ra basis functions with spatial angular momentum greater than 5 ($i$, $k$ etc.) 
and the neglect of quantum electrodynamic corrections \cite{Skripnikov:2021} 
lead to a comparable errors having the same sign and resulting in a systematic overestimation of three lowest excitation energies of free Ra$^{+}$ by 220-260 cm$^{-1}$; one could expect that the corresponding correction of calculated transition energies for the complex ions would rectify the estimates 
(see also \cite{Zaitsevskii:2022}).  

\begin{figure}[htp]
\begin{center}
\includegraphics[width=0.7\columnwidth]{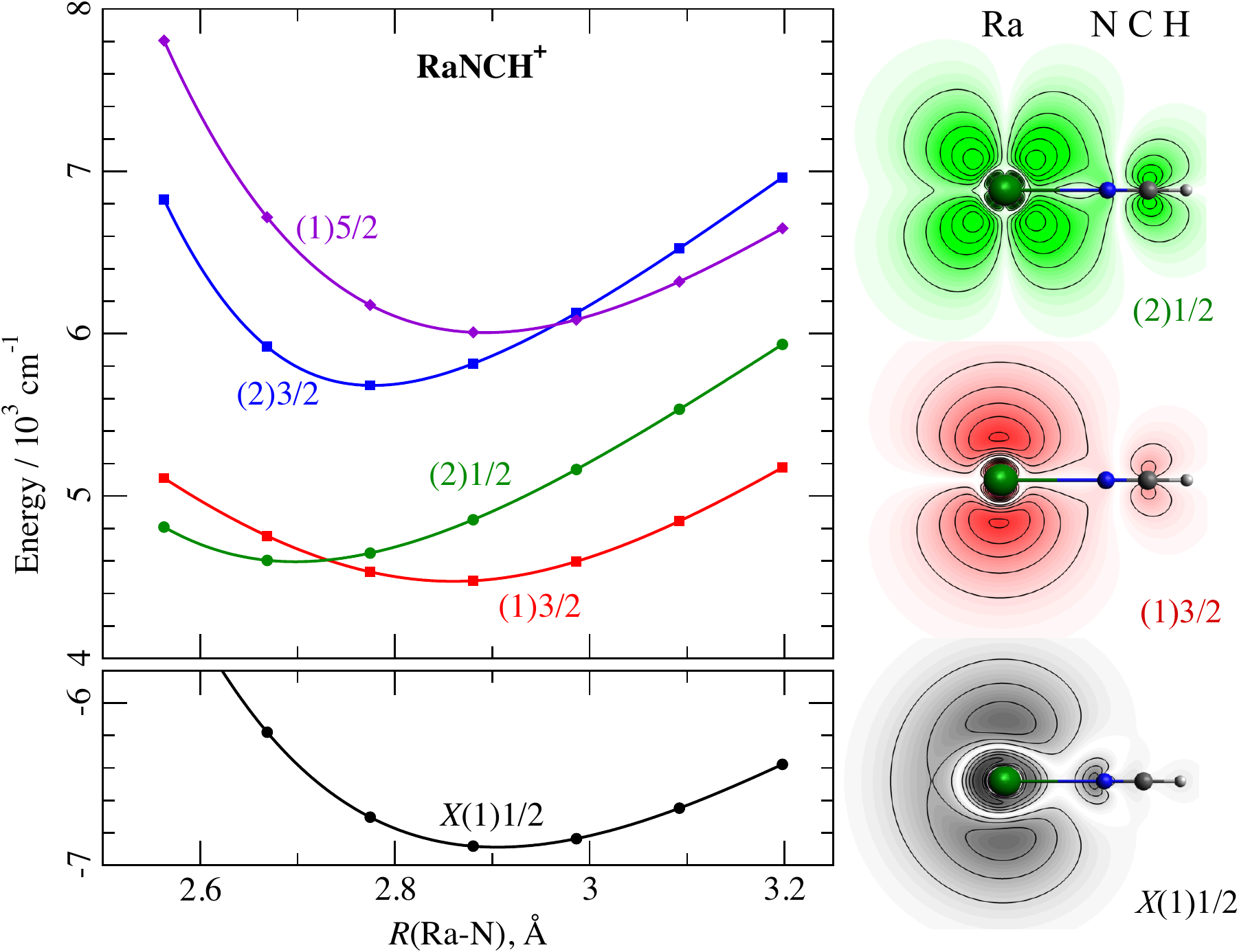}
\caption{
FS RCCSD / RCCSD(T) energies for five lowest electronic states of linear RaNCH$^+$ as functions of the Ra--N internuclear separation (left; the zero energy corresponds to the ground-state dissociation limit)
and absolute values for unpaired-electron natural spinors for the ground ($X(1)1/2$) and low-energy excited  (1)3/2 and (2)1/2 of along the symmetry plane (right; isoline values are separated by 0.02
$a_0^{-3/2}$).
\label{ranchstretch}
}
\end{center}
\end{figure}

For the RaNCH$^+$ complex we perform a study on the dependence of FS RCC energies on the Ra-N internuclear separation and Ra-N-C valence angle, leaving the ligand geometric parameters at their RDFT equilibrium values in the complex. In the case of stretching deformation, the results 
can be
affected by basis set superposition errors (BSSE) and the neglect of higher excitation contributions in the cluster operator. To reduce the corresponding errors, we evaluate the excited state energies by adding the geometry-dependent FS RCCSD excitation energies to the ground-state potential calculated by the single-reference RCC method with perturbative account of contributions from triple excitations (RCCSD(T)) and including the counterpoise BSSE corrections (ref. \cite{Isaev:21,Pazyuk:15}). The resulting potential functions for lowest electronic states and equilibrium parameters are displayed in Fig.~\ref{ranchstretch} and Table~\ref{spectro}, respectively.       
 
According to the calculations with ground-state equilibrium Ra--N separation and ligand geometry, the complex in all excited states under study should be linear.

To estimate prospects for laser-coolability of the proposed compounds 
we calculated FC-factors for vibronic transitions 
between ground and five lowest-lying excited electronic states for RaNCH$^+$ cation
as we would expect that chemical bonding situation is quite similar to this in other compounds.
 We accounted only for stretching mode 
between Ra and ligand, thus effectively reducing the vibrational problem to this in diatomic pseudomolecule RaX, where 
the
pseudoatom
X has atomic mass equal to mass of NCH.  The calculated points on PEC
are approximated by Morse potential and then FC factors together with equilibrium distance $R_e$ and vibrational
quanta $\omega_{\rm stretch}$ are obtained from the parameters of corresponding approximation.
The results are provided in the Table~\ref{fc-ranch}. It can be seen that vibronic transitions from 
the ground vibrational state of the state 1(3/2), 2(3/2) and
3(1/2) to the first three vibrational levels $X$  state posses a sum of the greatest FC-factors 
sufficiently
close to unity 
to be 
prospective for
using in quasiclosed optical cycling loop. 
At the same time, for vibronic transitions between 2(1/2) and ground electronic states
FC-matrix is clearly non-diagonal, and thus effective cooling loop for these transitions cannot be established. 

An insight into the origin of different dependences of potential energies for different low-lying states on the Ra-N separation can be gained from an
one-electron picture accounting for correlation effects. Such picture can be obtained via transforming the model-space projection of $(0h1p)$-sector FS RCC wavefunctions to a single-determinant form. Unpaired-electron spinors provide approximations for natural spinors (NS) of the corresponding states, one spinor per state. To better describe the shape of these spinors, the FS RCC solutions were obtained in larger model spaces (up to 50 active spinors) using the determinant-shifting technique to suppress the effect of possible intruder states \cite{Zaitsevskii:17}; 
the energy estimates are practically insensitive to this extension.
Fig.~\ref{ranchstretch} provides the plots of absolute values of model natural spinors (square roots of approximate unpaired electron densities) for several states. 

Non-diagonality of FC-matrix is related to a somewhat different nature of chemical  bonding in the ground and 
low-lying excited states.
In Fig.\ref{ranchstretch} one can see that the ground-state unpaired electron spinor is 
strongly locaized on the radium atom
and nearly vanishes in the vicinity of a surface separating Ra and N domains.
In other words, its shape is typical for a non-bonding spinor
of the first type, according to classification of non-bonding MOs/spinors in \cite{Isaev:10a},
and the contribution of the unpaired electron to the Ra-N bonding
is negligible. The corresponding spinors for the
first excited (1)3/2 state is slightly delocalized between the Ra ion and the ligand; the delocalization is more pronounced for the next (2)1/2 state.
For both excited states, these spinors pass through zero between the nitrogen and carbon atoms and thus can be interpreted
in MO-LCAO approach
in terms of admixture of the unoccupied anti-bonding spinor of the N-C fragment to Ra$^+$ spinors. The unpaired electron density in the area associated with the Ra-N bond is non-negligible for the (1)3/2 state and more significant for the (2)1/2. This picture indicates certain contributions of unpaired electron to Ra-N bond strengthening in these states, quite weak for (1)3/2 and significant for (2)1/2 and agrees well with the decrease of equilibrium separations between Ra and N, rather small for the former state and more significant for the latter one (Table \ref{spectro}).   

\begin{table}[h!]\caption{ RCCSD(T)/FS RCCSD equilibrium parameters of low-lying states of RaNCH$^+$ computed with the frozen ligand geometry. 
\label{spectro}}
\begin{center}
\begin{tabular}{lcccccc}
\hline \\ 
 State              &  $R_e$, \AA & $T_{e}$, $10^{3}$ cm$^{-1}$  &$\omega_{\rm stretch}$, cm$^{-1}$ \\ 
\\
   ($X$)1/2         &  2.905    &   0    &  161 \\    
   (1)3/2          &   2.860     & 11.36 & 136 \\  
   (2)1/2          &   2.698     & 11.48 & 155 \\    
   (2)3/2          &   2.777     & 12.57 & 207 \\   
   (3)1/2          &   3.011     & 15.99 & 144\\   
\\  \hline
\end{tabular}
\end{center}
\end{table} 

\begin{table}[h!]\caption{ Franck-Condon factors for vibronic transitions between ground vibrational state (0') of the
excited electronic states and three lowest vibrational states (0, 1 and 2) of the ground electronic state $X$.\label{fc-ranch}
FC-factors for transition from (1)5/2 are not provided due to zeroing of E1 electronic transition matrix element.}
\begin{center}
\begin{tabular}{l|cccccc}
 State                   &  (1)3/2    &  (2)1/2 & (2)3/2 & (1)5/2   &   (3)1/2 \\ 
\hline
   $0'\rightarrow 0$   &0.830 & 0.078     & 0.364& -- & 0.617\\    
   $0'\rightarrow 1$   & 0.127  & 0.151     &0.348  & --  & 0.315\\  
   $0'\rightarrow 2$   & 0.030  &  0.178   & 0.186 & --  & 0.062\\  
   \hline  
   Sum                     & 0.987  &  0.407    &0.898  & -- &0.994 \\   
  
\end{tabular}
\end{center}
\end{table} 
 \clearpage
\section{Conclusion}
We analyse electronic structure of a series of
Ra-containing molecular cations from the point of laser-coolability. Such cations can be used for a number of research, including searches for \podd and \todd violating interactions in molecules. 
The optical spectroscopic properties of all the considered molecular cations can be basically described in the framework 
of simple physical picture as attachment of Ra$^+$ to light-element ligand. This point is supported by that  
the electronic transition spectra of complexes closely remind weakly perturbed Ra+ spectra, 
that the binding energies are very moderate and inferior even to the lowest excitation energies and 
the equilibrium symmetry of considered complexes
is as high as symmetry of free ligand. Taking
 RaNCH$^+$ as example we calculate molecular parameters of the ground and five lowest-lying excited electronic states and 
 estimate prospects for creation of optical cycling center in this molecular cation.
We found that the excited states (1)3/2, (2)3/2 and (3)1/2 
of RaNCH$^+$ (or possibly analogous states of some similar systems) can
have sufficiently ``parallel'' surfaces of potential energy to the ground
electronic state (and, thus, almost-diagonal FC-matrix for vibronic transitions between these states)
to organise quasiclosed optical cycle, while ``parallelism'' of the state (2) 1/2 is 
fully destroyed  by a partial unpaired electron transfer to low-lying empty spinors of ligand, thus strengthening the chemical bond and displacing the equilibrium  
Ra$-$ligand distance.
\section{Acknowledgment}
 The authors are very grateful to C. Zuelch, K. Gaul and R. Berger for fruitful discussions in 2021.
The calculations have been carried out using computing resources of the federal collective usage center Complex for Simulation and Data Processing for Mega-science Facilities at National Research Centre ``Kurchatov Institute'', http://ckp.nrcki.ru/.
The work is supported by RSF grant No. 21-42-04411. 



 \bibliographystyle{apsrev4-2}
 \bibliography{/Users/Timur/iCloudDrive/AKbib/az}

\begin{thebibliography}{40}%
\makeatletter
\providecommand \@ifxundefined [1]{%
 \@ifx{#1\undefined}
}%
\providecommand \@ifnum [1]{%
 \ifnum #1\expandafter \@firstoftwo
 \else \expandafter \@secondoftwo
 \fi
}%
\providecommand \@ifx [1]{%
 \ifx #1\expandafter \@firstoftwo
 \else \expandafter \@secondoftwo
 \fi
}%
\providecommand \natexlab [1]{#1}%
\providecommand \enquote  [1]{``#1''}%
\providecommand \bibnamefont  [1]{#1}%
\providecommand \bibfnamefont [1]{#1}%
\providecommand \citenamefont [1]{#1}%
\providecommand \href@noop [0]{\@secondoftwo}%
\providecommand \href [0]{\begingroup \@sanitize@url \@href}%
\providecommand \@href[1]{\@@startlink{#1}\@@href}%
\providecommand \@@href[1]{\endgroup#1\@@endlink}%
\providecommand \@sanitize@url [0]{\catcode `\\12\catcode `\$12\catcode
  `\&12\catcode `\#12\catcode `\^12\catcode `\_12\catcode `\%12\relax}%
\providecommand \@@startlink[1]{}%
\providecommand \@@endlink[0]{}%
\providecommand \url  [0]{\begingroup\@sanitize@url \@url }%
\providecommand \@url [1]{\endgroup\@href {#1}{\urlprefix }}%
\providecommand \urlprefix  [0]{URL }%
\providecommand \Eprint [0]{\href }%
\providecommand \doibase [0]{https://doi.org/}%
\providecommand \selectlanguage [0]{\@gobble}%
\providecommand \bibinfo  [0]{\@secondoftwo}%
\providecommand \bibfield  [0]{\@secondoftwo}%
\providecommand \translation [1]{[#1]}%
\providecommand \BibitemOpen [0]{}%
\providecommand \bibitemStop [0]{}%
\providecommand \bibitemNoStop [0]{.\EOS\space}%
\providecommand \EOS [0]{\spacefactor3000\relax}%
\providecommand \BibitemShut  [1]{\csname bibitem#1\endcsname}%
\let\auto@bib@innerbib\@empty
\bibitem [{\citenamefont {Flambaum}\ and\ \citenamefont
  {Khriplovich}(1985)}]{flambaum:1985}%
  \BibitemOpen
  \bibfield  {author} {\bibinfo {author} {\bibfnamefont {V.~V.}\ \bibnamefont
  {Flambaum}}\ and\ \bibinfo {author} {\bibfnamefont {I.~B.}\ \bibnamefont
  {Khriplovich}},\ }\href@noop {} {\bibfield  {journal} {\bibinfo  {journal}
  {Phys. Lett. A}\ }\textbf {\bibinfo {volume} {110}},\ \bibinfo {pages} {121}
  (\bibinfo {year} {1985})}\BibitemShut {NoStop}%
\bibitem [{\citenamefont {Kozlov}\ and\ \citenamefont
  {Labzowsky}(1995)}]{kozlov:1995}%
  \BibitemOpen
  \bibfield  {author} {\bibinfo {author} {\bibfnamefont {M.~G.}\ \bibnamefont
  {Kozlov}}\ and\ \bibinfo {author} {\bibfnamefont {L.~N.}\ \bibnamefont
  {Labzowsky}},\ }\href {https://doi.org/10.1088/0953-4075/28/10/008}
  {\bibfield  {journal} {\bibinfo  {journal} {J. Phys. B}\ }\textbf {\bibinfo
  {volume} {28}},\ \bibinfo {pages} {1933} (\bibinfo {year}
  {1995})}\BibitemShut {NoStop}%
\bibitem [{\citenamefont {Berger}(2004)}]{berger:2004a}%
  \BibitemOpen
  \bibfield  {author} {\bibinfo {author} {\bibfnamefont {R.}~\bibnamefont
  {Berger}},\ }in\ \href@noop {} {\emph {\bibinfo {booktitle} {Relativistic
  Electronic Structure Theory, Part: 2, Applications}}},\ \bibinfo {editor}
  {edited by\ \bibinfo {editor} {\bibfnamefont {P.}~\bibnamefont
  {Schwerdtfeger}}}\ (\bibinfo  {publisher} {Elsevier},\ \bibinfo {address}
  {Netherlands},\ \bibinfo {year} {2004})\ Chap.~\bibinfo {chapter} {4}, pp.\
  \bibinfo {pages} {188--288}\BibitemShut {NoStop}%
\bibitem [{\citenamefont {Vutha}\ \emph {et~al.}(2011)\citenamefont {Vutha},
  \citenamefont {Spaun}, \citenamefont {Gurevich}, \citenamefont {Hutzler},
  \citenamefont {Kirilov}, \citenamefont {Doyle}, \citenamefont {Gabrielse},\
  and\ \citenamefont {DeMille}}]{acme:11}%
  \BibitemOpen
  \bibfield  {author} {\bibinfo {author} {\bibfnamefont {A.~C.}\ \bibnamefont
  {Vutha}}, \bibinfo {author} {\bibfnamefont {B.}~\bibnamefont {Spaun}},
  \bibinfo {author} {\bibfnamefont {Y.~V.}\ \bibnamefont {Gurevich}}, \bibinfo
  {author} {\bibfnamefont {N.~R.}\ \bibnamefont {Hutzler}}, \bibinfo {author}
  {\bibfnamefont {E.}~\bibnamefont {Kirilov}}, \bibinfo {author} {\bibfnamefont
  {J.~M.}\ \bibnamefont {Doyle}}, \bibinfo {author} {\bibfnamefont
  {G.}~\bibnamefont {Gabrielse}},\ and\ \bibinfo {author} {\bibfnamefont
  {D.}~\bibnamefont {DeMille}},\ }\href
  {https://doi.org/{10.1103/PhysRevA.84.034502}} {\bibfield  {journal}
  {\bibinfo  {journal} {{Physical Review A}}\ }\textbf {\bibinfo {volume}
  {{84}}},\ \bibinfo {pages} {034502} (\bibinfo {year} {{2011}})}\BibitemShut
  {NoStop}%
\bibitem [{\citenamefont {Kozyryev}\ and\ \citenamefont
  {Hutzler}(2017)}]{Kozyryev:2017b}%
  \BibitemOpen
  \bibfield  {author} {\bibinfo {author} {\bibfnamefont {I.}~\bibnamefont
  {Kozyryev}}\ and\ \bibinfo {author} {\bibfnamefont {N.}~\bibnamefont
  {Hutzler}},\ }\href {https://arxiv.org/abs/1705.11020} {\bibfield  {journal}
  {\bibinfo  {journal} {Phys. Rev. Lett.}\ }\textbf {\bibinfo {volume} {119}},\
  \bibinfo {pages} {133002} (\bibinfo {year} {2017})}\BibitemShut {NoStop}%
\bibitem [{\citenamefont {Isaev}\ \emph {et~al.}(2017)\citenamefont {Isaev},
  \citenamefont {Zaitsevskii},\ and\ \citenamefont {Eliav}}]{Isaev:17}%
  \BibitemOpen
  \bibfield  {author} {\bibinfo {author} {\bibfnamefont {T.~A.}\ \bibnamefont
  {Isaev}}, \bibinfo {author} {\bibfnamefont {A.~V.}\ \bibnamefont
  {Zaitsevskii}},\ and\ \bibinfo {author} {\bibfnamefont {E.}~\bibnamefont
  {Eliav}},\ }\href {https://doi.org/10.1088/1361-6455/aa8f34} {\bibfield
  {journal} {\bibinfo  {journal} {J.\ Phys.\ B:\ At.\ Mol.\ Opt.\ Phys.}\
  }\textbf {\bibinfo {volume} {50}},\ \bibinfo {pages} {225101} (\bibinfo
  {year} {2017})}\BibitemShut {NoStop}%
\bibitem [{\citenamefont {Gaul}\ and\ \citenamefont
  {Berger}(2017)}]{Gaul:2017}%
  \BibitemOpen
  \bibfield  {author} {\bibinfo {author} {\bibfnamefont {K.}~\bibnamefont
  {Gaul}}\ and\ \bibinfo {author} {\bibfnamefont {R.}~\bibnamefont {Berger}},\
  }\href {https://doi.org/10.1063/1.4985567} {\bibfield  {journal} {\bibinfo
  {journal} {The Journal of Chemical Physics}\ }\textbf {\bibinfo {volume}
  {147}},\ \bibinfo {pages} {014109} (\bibinfo {year} {2017})}\BibitemShut
  {NoStop}%
\bibitem [{\citenamefont {Auerbach}\ \emph {et~al.}(1996)\citenamefont
  {Auerbach}, \citenamefont {Flambaum},\ and\ \citenamefont
  {Spevak}}]{Auerbach:96}%
  \BibitemOpen
  \bibfield  {author} {\bibinfo {author} {\bibfnamefont {N.}~\bibnamefont
  {Auerbach}}, \bibinfo {author} {\bibfnamefont {V.~V.}\ \bibnamefont
  {Flambaum}},\ and\ \bibinfo {author} {\bibfnamefont {V.}~\bibnamefont
  {Spevak}},\ }\href@noop {} {\bibfield  {journal} {\bibinfo  {journal} {Phys.
  Rev. Lett.}\ }\textbf {\bibinfo {volume} {76}},\ \bibinfo {pages} {4316}
  (\bibinfo {year} {1996})}\BibitemShut {NoStop}%
\bibitem [{\citenamefont {Spevak}\ \emph {et~al.}(1997)\citenamefont {Spevak},
  \citenamefont {Auerbach},\ and\ \citenamefont {Flambaum}}]{Spevak:97}%
  \BibitemOpen
  \bibfield  {author} {\bibinfo {author} {\bibfnamefont {V.}~\bibnamefont
  {Spevak}}, \bibinfo {author} {\bibfnamefont {N.}~\bibnamefont {Auerbach}},\
  and\ \bibinfo {author} {\bibfnamefont {V.~V.}\ \bibnamefont {Flambaum}},\
  }\href {https://doi.org/{10.1103/PhysRevC.56.1357}} {\bibfield  {journal}
  {\bibinfo  {journal} {Phys. Rev. C}\ }\textbf {\bibinfo {volume} {56}},\
  \bibinfo {pages} {1357} (\bibinfo {year} {1997})}\BibitemShut {NoStop}%
\bibitem [{\citenamefont {Garcia~Ruiz}\ \emph {et~al.}(2020)\citenamefont
  {Garcia~Ruiz}, \citenamefont {Berger}, \citenamefont {Billowes},
  \citenamefont {Binnersley}, \citenamefont {Bissell}, \citenamefont {Breier},
  \citenamefont {Brinson}, \citenamefont {Chrysalidis}, \citenamefont
  {Cocolios}, \citenamefont {Cooper}, \citenamefont {Flanagan}, \citenamefont
  {Giesen}, \citenamefont {de~Groote}, \citenamefont {Franchoo}, \citenamefont
  {Gustafsson}, \citenamefont {Isaev}, \citenamefont {Koszor{\'u}s},
  \citenamefont {Neyens}, \citenamefont {Perrett}, \citenamefont {Ricketts},
  \citenamefont {Rothe}, \citenamefont {Schweikhard}, \citenamefont {Vernon},
  \citenamefont {Wendt}, \citenamefont {Wienholtz}, \citenamefont {Wilkins},\
  and\ \citenamefont {Yang}}]{GarciaRuiz:2020}%
  \BibitemOpen
  \bibfield  {author} {\bibinfo {author} {\bibfnamefont {R.~F.}\ \bibnamefont
  {Garcia~Ruiz}}, \bibinfo {author} {\bibfnamefont {R.}~\bibnamefont {Berger}},
  \bibinfo {author} {\bibfnamefont {J.}~\bibnamefont {Billowes}}, \bibinfo
  {author} {\bibfnamefont {C.~L.}\ \bibnamefont {Binnersley}}, \bibinfo
  {author} {\bibfnamefont {M.~L.}\ \bibnamefont {Bissell}}, \bibinfo {author}
  {\bibfnamefont {A.~A.}\ \bibnamefont {Breier}}, \bibinfo {author}
  {\bibfnamefont {A.~J.}\ \bibnamefont {Brinson}}, \bibinfo {author}
  {\bibfnamefont {K.}~\bibnamefont {Chrysalidis}}, \bibinfo {author}
  {\bibfnamefont {T.~E.}\ \bibnamefont {Cocolios}}, \bibinfo {author}
  {\bibfnamefont {B.~S.}\ \bibnamefont {Cooper}}, \bibinfo {author}
  {\bibfnamefont {K.~T.}\ \bibnamefont {Flanagan}}, \bibinfo {author}
  {\bibfnamefont {T.~F.}\ \bibnamefont {Giesen}}, \bibinfo {author}
  {\bibfnamefont {R.~P.}\ \bibnamefont {de~Groote}}, \bibinfo {author}
  {\bibfnamefont {S.}~\bibnamefont {Franchoo}}, \bibinfo {author}
  {\bibfnamefont {F.~P.}\ \bibnamefont {Gustafsson}}, \bibinfo {author}
  {\bibfnamefont {T.~A.}\ \bibnamefont {Isaev}}, \bibinfo {author}
  {\bibfnamefont {{\'A}.}~\bibnamefont {Koszor{\'u}s}}, \bibinfo {author}
  {\bibfnamefont {G.}~\bibnamefont {Neyens}}, \bibinfo {author} {\bibfnamefont
  {H.~A.}\ \bibnamefont {Perrett}}, \bibinfo {author} {\bibfnamefont {C.~M.}\
  \bibnamefont {Ricketts}}, \bibinfo {author} {\bibfnamefont {S.}~\bibnamefont
  {Rothe}}, \bibinfo {author} {\bibfnamefont {L.}~\bibnamefont {Schweikhard}},
  \bibinfo {author} {\bibfnamefont {A.~R.}\ \bibnamefont {Vernon}}, \bibinfo
  {author} {\bibfnamefont {K.~D.~A.}\ \bibnamefont {Wendt}}, \bibinfo {author}
  {\bibfnamefont {F.}~\bibnamefont {Wienholtz}}, \bibinfo {author}
  {\bibfnamefont {S.~G.}\ \bibnamefont {Wilkins}},\ and\ \bibinfo {author}
  {\bibfnamefont {X.~F.}\ \bibnamefont {Yang}},\ }\href
  {https://doi.org/10.1038/s41586-020-2299-4} {\bibfield  {journal} {\bibinfo
  {journal} {Nature}\ }\textbf {\bibinfo {volume} {581}},\ \bibinfo {pages}
  {396} (\bibinfo {year} {2020})}\BibitemShut {NoStop}%
\bibitem [{\citenamefont {Isaev}\ \emph {et~al.}(2010)\citenamefont {Isaev},
  \citenamefont {Hoekstra},\ and\ \citenamefont {Berger}}]{Isaev:10a}%
  \BibitemOpen
  \bibfield  {author} {\bibinfo {author} {\bibfnamefont {T.~A.}\ \bibnamefont
  {Isaev}}, \bibinfo {author} {\bibfnamefont {S.}~\bibnamefont {Hoekstra}},\
  and\ \bibinfo {author} {\bibfnamefont {R.}~\bibnamefont {Berger}},\
  }\href@noop {} {\bibfield  {journal} {\bibinfo  {journal} {Phys. Rev. A}\
  }\textbf {\bibinfo {volume} {82}},\ \bibinfo {pages} {052521} (\bibinfo
  {year} {2010})}\BibitemShut {NoStop}%
\bibitem [{\citenamefont {{Isaev}}\ and\ \citenamefont
  {{Berger}}(2013)}]{Isaev:13}%
  \BibitemOpen
  \bibfield  {author} {\bibinfo {author} {\bibfnamefont {T.~A.}\ \bibnamefont
  {{Isaev}}}\ and\ \bibinfo {author} {\bibfnamefont {R.}~\bibnamefont
  {{Berger}}},\ }\href@noop {} {\bibfield  {journal} {\bibinfo  {journal}
  {ArXiv e-prints {physics.chem-ph:1302.5682}}\ } (\bibinfo {year} {2013})},\
  \Eprint {https://arxiv.org/abs/1302.5682} {arXiv:1302.5682 [physics.chem-ph]}
  \BibitemShut {NoStop}%
\bibitem [{\citenamefont {Zaitsevskii}\ \emph {et~al.}(2022)\citenamefont
  {Zaitsevskii}, \citenamefont {Skripnikov}, \citenamefont {Mosyagin},
  \citenamefont {Isaev}, \citenamefont {Berger}, \citenamefont {Breier},\ and\
  \citenamefont {Giesen}}]{Zaitsevskii:2022}%
  \BibitemOpen
  \bibfield  {author} {\bibinfo {author} {\bibfnamefont {A.}~\bibnamefont
  {Zaitsevskii}}, \bibinfo {author} {\bibfnamefont {L.~V.}\ \bibnamefont
  {Skripnikov}}, \bibinfo {author} {\bibfnamefont {N.~S.}\ \bibnamefont
  {Mosyagin}}, \bibinfo {author} {\bibfnamefont {T.}~\bibnamefont {Isaev}},
  \bibinfo {author} {\bibfnamefont {R.}~\bibnamefont {Berger}}, \bibinfo
  {author} {\bibfnamefont {A.~A.}\ \bibnamefont {Breier}},\ and\ \bibinfo
  {author} {\bibfnamefont {T.~F.}\ \bibnamefont {Giesen}},\ }\href
  {https://doi.org/10.1063/5.0079618} {\bibfield  {journal} {\bibinfo
  {journal} {The Journal of Chemical Physics}\ }\textbf {\bibinfo {volume}
  {156}},\ \bibinfo {pages} {044306} (\bibinfo {year} {2022})}\BibitemShut
  {NoStop}%
\bibitem [{\citenamefont {Lien}\ \emph {et~al.}(2011)\citenamefont {Lien},
  \citenamefont {Williams},\ and\ \citenamefont {Odom}}]{Lien:2011}%
  \BibitemOpen
  \bibfield  {author} {\bibinfo {author} {\bibfnamefont {C.-Y.}\ \bibnamefont
  {Lien}}, \bibinfo {author} {\bibfnamefont {S.~R.}\ \bibnamefont {Williams}},\
  and\ \bibinfo {author} {\bibfnamefont {B.}~\bibnamefont {Odom}},\ }\href
  {https://doi.org/10.1039/c1cp21201j} {\bibfield  {journal} {\bibinfo
  {journal} {Physical Chemistry Chemical Physics}\ }\textbf {\bibinfo {volume}
  {13}},\ \bibinfo {pages} {18825} (\bibinfo {year} {2011})}\BibitemShut
  {NoStop}%
\bibitem [{\citenamefont {Lien}\ \emph {et~al.}(2014)\citenamefont {Lien},
  \citenamefont {Seck}, \citenamefont {Lin}, \citenamefont {Nguyen},
  \citenamefont {Tabor},\ and\ \citenamefont {Odom}}]{Lien:2014}%
  \BibitemOpen
  \bibfield  {author} {\bibinfo {author} {\bibfnamefont {C.-Y.}\ \bibnamefont
  {Lien}}, \bibinfo {author} {\bibfnamefont {C.~M.}\ \bibnamefont {Seck}},
  \bibinfo {author} {\bibfnamefont {Y.-W.}\ \bibnamefont {Lin}}, \bibinfo
  {author} {\bibfnamefont {J.~H.}\ \bibnamefont {Nguyen}}, \bibinfo {author}
  {\bibfnamefont {D.~A.}\ \bibnamefont {Tabor}},\ and\ \bibinfo {author}
  {\bibfnamefont {B.~C.}\ \bibnamefont {Odom}},\ }\bibfield  {journal}
  {\bibinfo  {journal} {Nature Communications}\ }\textbf {\bibinfo {volume}
  {5}},\ \href {https://doi.org/10.1038/ncomms5783} {10.1038/ncomms5783}
  (\bibinfo {year} {2014})\BibitemShut {NoStop}%
\bibitem [{\citenamefont {Stollenwerk}\ \emph {et~al.}(2020)\citenamefont
  {Stollenwerk}, \citenamefont {Antonov}, \citenamefont {Venkataramanababu},
  \citenamefont {Lin},\ and\ \citenamefont {Odom}}]{Stollenwerk:2020}%
  \BibitemOpen
  \bibfield  {author} {\bibinfo {author} {\bibfnamefont {P.~R.}\ \bibnamefont
  {Stollenwerk}}, \bibinfo {author} {\bibfnamefont {I.~O.}\ \bibnamefont
  {Antonov}}, \bibinfo {author} {\bibfnamefont {S.}~\bibnamefont
  {Venkataramanababu}}, \bibinfo {author} {\bibfnamefont {Y.-W.}\ \bibnamefont
  {Lin}},\ and\ \bibinfo {author} {\bibfnamefont {B.~C.}\ \bibnamefont
  {Odom}},\ }\href {https://doi.org/10.1103/physrevlett.125.113201} {\bibfield
  {journal} {\bibinfo  {journal} {Physical Review Letters}\ }\textbf {\bibinfo
  {volume} {125}},\ \bibinfo {pages} {113201} (\bibinfo {year}
  {2020})}\BibitemShut {NoStop}%
\bibitem [{\citenamefont {Cairncross}\ \emph {et~al.}(2017)\citenamefont
  {Cairncross}, \citenamefont {Gresh}, \citenamefont {Grau}, \citenamefont
  {Cossel}, \citenamefont {Roussy}, \citenamefont {Ni}, \citenamefont {Zhou},
  \citenamefont {Ye},\ and\ \citenamefont {Cornell}}]{Cairncross:2017}%
  \BibitemOpen
  \bibfield  {author} {\bibinfo {author} {\bibfnamefont {W.~B.}\ \bibnamefont
  {Cairncross}}, \bibinfo {author} {\bibfnamefont {D.~N.}\ \bibnamefont
  {Gresh}}, \bibinfo {author} {\bibfnamefont {M.}~\bibnamefont {Grau}},
  \bibinfo {author} {\bibfnamefont {K.~C.}\ \bibnamefont {Cossel}}, \bibinfo
  {author} {\bibfnamefont {T.~S.}\ \bibnamefont {Roussy}}, \bibinfo {author}
  {\bibfnamefont {Y.}~\bibnamefont {Ni}}, \bibinfo {author} {\bibfnamefont
  {Y.}~\bibnamefont {Zhou}}, \bibinfo {author} {\bibfnamefont {J.}~\bibnamefont
  {Ye}},\ and\ \bibinfo {author} {\bibfnamefont {E.~A.}\ \bibnamefont
  {Cornell}},\ }\href
  {https://doi.org/https://doi.org/10.1103/PhysRevLett.119.153001} {\bibfield
  {journal} {\bibinfo  {journal} {Physical Review Letters}\ }\textbf {\bibinfo
  {volume} {119}},\ \bibinfo {pages} {153001} (\bibinfo {year}
  {2017})}\BibitemShut {NoStop}%
\bibitem [{\citenamefont {Zakharova}\ and\ \citenamefont
  {Petrov}(2021)}]{Zakharova:2021}%
  \BibitemOpen
  \bibfield  {author} {\bibinfo {author} {\bibfnamefont {A.}~\bibnamefont
  {Zakharova}}\ and\ \bibinfo {author} {\bibfnamefont {A.}~\bibnamefont
  {Petrov}},\ }\href {https://doi.org/10.1103/physreva.103.032819} {\bibfield
  {journal} {\bibinfo  {journal} {Physical Review A}\ }\textbf {\bibinfo
  {volume} {103}},\ \bibinfo {pages} {032819} (\bibinfo {year}
  {2021})}\BibitemShut {NoStop}%
\bibitem [{\citenamefont {Zakharova}(2022)}]{Zakharova:2022}%
  \BibitemOpen
  \bibfield  {author} {\bibinfo {author} {\bibfnamefont {A.}~\bibnamefont
  {Zakharova}},\ }\href {https://doi.org/10.1103/physreva.105.032811}
  {\bibfield  {journal} {\bibinfo  {journal} {Physical Review A}\ }\textbf
  {\bibinfo {volume} {105}},\ \bibinfo {pages} {032811} (\bibinfo {year}
  {2022})}\BibitemShut {NoStop}%
\bibitem [{\citenamefont {Fan}\ \emph {et~al.}(2021)\citenamefont {Fan},
  \citenamefont {Holliman}, \citenamefont {Shi}, \citenamefont {Zhang},
  \citenamefont {Straus}, \citenamefont {Li}, \citenamefont {Buechele},\ and\
  \citenamefont {Jayich}}]{Fan:2021}%
  \BibitemOpen
  \bibfield  {author} {\bibinfo {author} {\bibfnamefont {M.}~\bibnamefont
  {Fan}}, \bibinfo {author} {\bibfnamefont {C.}~\bibnamefont {Holliman}},
  \bibinfo {author} {\bibfnamefont {X.}~\bibnamefont {Shi}}, \bibinfo {author}
  {\bibfnamefont {H.}~\bibnamefont {Zhang}}, \bibinfo {author} {\bibfnamefont
  {M.}~\bibnamefont {Straus}}, \bibinfo {author} {\bibfnamefont
  {X.}~\bibnamefont {Li}}, \bibinfo {author} {\bibfnamefont {S.}~\bibnamefont
  {Buechele}},\ and\ \bibinfo {author} {\bibfnamefont {A.}~\bibnamefont
  {Jayich}},\ }\href {https://doi.org/10.1103/physrevlett.126.023002}
  {\bibfield  {journal} {\bibinfo  {journal} {Physical Review Letters}\
  }\textbf {\bibinfo {volume} {126}},\ \bibinfo {pages} {023002} (\bibinfo
  {year} {2021})}\BibitemShut {NoStop}%
\bibitem [{\citenamefont {DiRosa}(2004)}]{DiRosa:2004}%
  \BibitemOpen
  \bibfield  {author} {\bibinfo {author} {\bibfnamefont {M.~D.}\ \bibnamefont
  {DiRosa}},\ }\href@noop {} {\bibfield  {journal} {\bibinfo  {journal} {Eur.
  Phys. J. D}\ }\textbf {\bibinfo {volume} {31}},\ \bibinfo {pages} {395}
  (\bibinfo {year} {2004})}\BibitemShut {NoStop}%
\bibitem [{\citenamefont {Isaev}\ \emph {et~al.}(2021)\citenamefont {Isaev},
  \citenamefont {Zaitsevskii}, \citenamefont {Oleynichenko}, \citenamefont
  {Eliav}, \citenamefont {Breier}, \citenamefont {Giesen}, \citenamefont
  {{Garcia Ruiz}},\ and\ \citenamefont {Berger}}]{Isaev:21}%
  \BibitemOpen
  \bibfield  {author} {\bibinfo {author} {\bibfnamefont {T.~A.}\ \bibnamefont
  {Isaev}}, \bibinfo {author} {\bibfnamefont {A.~V.}\ \bibnamefont
  {Zaitsevskii}}, \bibinfo {author} {\bibfnamefont {A.}~\bibnamefont
  {Oleynichenko}}, \bibinfo {author} {\bibfnamefont {E.}~\bibnamefont {Eliav}},
  \bibinfo {author} {\bibfnamefont {A.~A.}\ \bibnamefont {Breier}}, \bibinfo
  {author} {\bibfnamefont {T.~F.}\ \bibnamefont {Giesen}}, \bibinfo {author}
  {\bibfnamefont {R.~F.}\ \bibnamefont {{Garcia Ruiz}}},\ and\ \bibinfo
  {author} {\bibfnamefont {R.}~\bibnamefont {Berger}},\ }\href
  {https://doi.org/10.1016/j.jqsrt.2021.107649} {\bibfield  {journal} {\bibinfo
   {journal} {{J}.\ {Q}uant.\ {S}pectrosc.\ {R}adiat.\ {Transfer}}\ }\textbf
  {\bibinfo {volume} {269}},\ \bibinfo {pages} {107649} (\bibinfo {year}
  {2021})}\BibitemShut {NoStop}%
\bibitem [{\citenamefont {Mosyagin}\ \emph {et~al.}(2010)\citenamefont
  {Mosyagin}, \citenamefont {Zaitsevskii},\ and\ \citenamefont
  {Titov}}]{Mosyagin:10a}%
  \BibitemOpen
  \bibfield  {author} {\bibinfo {author} {\bibfnamefont {N.~S.}\ \bibnamefont
  {Mosyagin}}, \bibinfo {author} {\bibfnamefont {A.}~\bibnamefont
  {Zaitsevskii}},\ and\ \bibinfo {author} {\bibfnamefont {A.~V.}\ \bibnamefont
  {Titov}},\ }\href
  {https://www.auburn.edu/cosam/departments/physics/iramp/1\_1/mosyagin\_zaitsevskii\_titov.pdf}
  {\bibfield  {journal} {\bibinfo  {journal} {{I}nt.\ {R}ev.\ {A}t.\ {M}ol.\
  {P}hys.}\ }\textbf {\bibinfo {volume} {1}},\ \bibinfo {pages} {63} (\bibinfo
  {year} {2010})}\BibitemShut {NoStop}%
\bibitem [{\citenamefont {Mosyagin}\ \emph {et~al.}(2016)\citenamefont
  {Mosyagin}, \citenamefont {Zaitsevskii}, \citenamefont {Skripnikov},\ and\
  \citenamefont {Titov}}]{Mosyagin:16}%
  \BibitemOpen
  \bibfield  {author} {\bibinfo {author} {\bibfnamefont {N.~S.}\ \bibnamefont
  {Mosyagin}}, \bibinfo {author} {\bibfnamefont {A.~V.}\ \bibnamefont
  {Zaitsevskii}}, \bibinfo {author} {\bibfnamefont {L.~V.}\ \bibnamefont
  {Skripnikov}},\ and\ \bibinfo {author} {\bibfnamefont {A.~V.}\ \bibnamefont
  {Titov}},\ }\href {https://doi.org/10.1002/qua.24978} {\bibfield  {journal}
  {\bibinfo  {journal} {{I}nt.\ {J}.\ {Q}uantum\ {C}hem.}\ }\textbf {\bibinfo
  {volume} {116}},\ \bibinfo {pages} {301} (\bibinfo {year}
  {2016})}\BibitemShut {NoStop}%
\bibitem [{\citenamefont {Mosyagin}\ and\ \citenamefont {Titov}()}]{ourGRECP}%
  \BibitemOpen
  \bibfield  {author} {\bibinfo {author} {\bibfnamefont {N.~S.}\ \bibnamefont
  {Mosyagin}}\ and\ \bibinfo {author} {\bibfnamefont {A.~V.}\ \bibnamefont
  {Titov}},\ }\href@noop {} {\bibinfo {title} {Generalized relativistic
  effective core potentials}},\ \bibinfo {note}
  {{h}ttp://www.qchem.pnpi.spb.ru/recp}\BibitemShut {NoStop}%
\bibitem [{\citenamefont {Mosyagin}\ \emph {et~al.}(2021)\citenamefont
  {Mosyagin}, \citenamefont {Oleynichenko}, \citenamefont {Zaitsevskii},
  \citenamefont {Kudrin}, \citenamefont {Pazyuk},\ and\ \citenamefont
  {Stolyarov}}]{Mosyagin:21}%
  \BibitemOpen
  \bibfield  {author} {\bibinfo {author} {\bibfnamefont {N.}~\bibnamefont
  {Mosyagin}}, \bibinfo {author} {\bibfnamefont {A.}~\bibnamefont
  {Oleynichenko}}, \bibinfo {author} {\bibfnamefont {A.}~\bibnamefont
  {Zaitsevskii}}, \bibinfo {author} {\bibfnamefont {A.}~\bibnamefont {Kudrin}},
  \bibinfo {author} {\bibfnamefont {E.}~\bibnamefont {Pazyuk}},\ and\ \bibinfo
  {author} {\bibfnamefont {A.}~\bibnamefont {Stolyarov}},\ }\href
  {https://doi.org/10.1016/j.jqsrt.2021.107532} {\bibfield  {journal} {\bibinfo
   {journal} {{J}.\ {Q}uant.\ {S}pectrosc.\ {R}adiat.\ {Transfer}}\ }\textbf
  {\bibinfo {volume} {263}},\ \bibinfo {pages} {107532} (\bibinfo {year}
  {2021})}\BibitemShut {NoStop}%
\bibitem [{\citenamefont {{van~W\"ullen}}(2010)}]{Wullen:10}%
  \BibitemOpen
  \bibfield  {author} {\bibinfo {author} {\bibfnamefont {C.}~\bibnamefont
  {{van~W\"ullen}}},\ }\href {https://doi.org/10.1524/zpch.2010.6114}
  {\bibfield  {journal} {\bibinfo  {journal} {{Z}.\ {P}hys.\ {C}hem.}\ }\textbf
  {\bibinfo {volume} {224}},\ \bibinfo {pages} {413} (\bibinfo {year}
  {2010})}\BibitemShut {NoStop}%
\bibitem [{\citenamefont {Adamo}\ and\ \citenamefont
  {Barone}(1999)}]{Adamo:99}%
  \BibitemOpen
  \bibfield  {author} {\bibinfo {author} {\bibfnamefont {C.}~\bibnamefont
  {Adamo}}\ and\ \bibinfo {author} {\bibfnamefont {V.}~\bibnamefont {Barone}},\
  }\href {https://doi.org/10.1063/1.478522} {\bibfield  {journal} {\bibinfo
  {journal} {J.\ Chem.\ Phys.}\ }\textbf {\bibinfo {volume} {110}},\ \bibinfo
  {pages} {6158} (\bibinfo {year} {1999})}\BibitemShut {NoStop}%
\bibitem [{\citenamefont {Sch\"{a}fer}\ \emph {et~al.}(1994)\citenamefont
  {Sch\"{a}fer}, \citenamefont {Huber},\ and\ \citenamefont
  {Ahlrichs}}]{Shaefer:94}%
  \BibitemOpen
  \bibfield  {author} {\bibinfo {author} {\bibfnamefont {A.}~\bibnamefont
  {Sch\"{a}fer}}, \bibinfo {author} {\bibfnamefont {C.}~\bibnamefont {Huber}},\
  and\ \bibinfo {author} {\bibfnamefont {R.}~\bibnamefont {Ahlrichs}},\ }\href
  {https://doi.org/10.1063/1.467146} {\bibfield  {journal} {\bibinfo  {journal}
  {{J}.\ {C}hem.\ {P}hys.}\ }\textbf {\bibinfo {volume} {100}},\ \bibinfo
  {pages} {5829} (\bibinfo {year} {1994})},\ \Eprint
  {https://arxiv.org/abs/https://doi.org/10.1063/1.467146}
  {https://doi.org/10.1063/1.467146} \BibitemShut {NoStop}%
\bibitem [{\citenamefont {Pritchard}\ \emph {et~al.}(2019)\citenamefont
  {Pritchard}, \citenamefont {Altarawy}, \citenamefont {Didier}, \citenamefont
  {Gibson},\ and\ \citenamefont {Windus}}]{Basissetexchange:19}%
  \BibitemOpen
  \bibfield  {author} {\bibinfo {author} {\bibfnamefont {B.~P.}\ \bibnamefont
  {Pritchard}}, \bibinfo {author} {\bibfnamefont {D.}~\bibnamefont {Altarawy}},
  \bibinfo {author} {\bibfnamefont {B.}~\bibnamefont {Didier}}, \bibinfo
  {author} {\bibfnamefont {T.~D.}\ \bibnamefont {Gibson}},\ and\ \bibinfo
  {author} {\bibfnamefont {T.~L.}\ \bibnamefont {Windus}},\ }\href
  {https://doi.org/10.1021/ci600510j} {\bibfield  {journal} {\bibinfo
  {journal} {{J}.\ {C}hem.\ {I}nf.\ {M}odel.}\ }\textbf {\bibinfo {volume}
  {47}},\ \bibinfo {pages} {1045} (\bibinfo {year} {2019})}\BibitemShut
  {NoStop}%
\bibitem [{\citenamefont {Visscher}\ \emph {et~al.}(2001)\citenamefont
  {Visscher}, \citenamefont {Eliav},\ and\ \citenamefont
  {Kaldor}}]{Visscher:01}%
  \BibitemOpen
  \bibfield  {author} {\bibinfo {author} {\bibfnamefont {L.}~\bibnamefont
  {Visscher}}, \bibinfo {author} {\bibfnamefont {E.}~\bibnamefont {Eliav}},\
  and\ \bibinfo {author} {\bibfnamefont {U.}~\bibnamefont {Kaldor}},\ }\href
  {https://doi.org/10.1063/1.1415746} {\bibfield  {journal} {\bibinfo
  {journal} {{J}.\ {C}hem.\ {P}hys.}\ }\textbf {\bibinfo {volume} {115}},\
  \bibinfo {pages} {9720} (\bibinfo {year} {2001})}\BibitemShut {NoStop}%
\bibitem [{\citenamefont {Dunning}(1989)}]{Dunning:89}%
  \BibitemOpen
  \bibfield  {author} {\bibinfo {author} {\bibfnamefont {T.~H.}\ \bibnamefont
  {Dunning}},\ }\href {https://doi.org/10.1063/1.456153} {\bibfield  {journal}
  {\bibinfo  {journal} {{J}.\ {C}hem.\ {P}hys.}\ }\textbf {\bibinfo {volume}
  {90}},\ \bibinfo {pages} {1007} (\bibinfo {year} {1989})}\BibitemShut
  {NoStop}%
\bibitem [{\citenamefont {Kendall}\ \emph {et~al.}(1992)\citenamefont
  {Kendall}, \citenamefont {{Dunning Jr.}},\ and\ \citenamefont
  {Harrison}}]{Kendall:92}%
  \BibitemOpen
  \bibfield  {author} {\bibinfo {author} {\bibfnamefont {R.~A.}\ \bibnamefont
  {Kendall}}, \bibinfo {author} {\bibfnamefont {T.~H.}\ \bibnamefont {{Dunning
  Jr.}}},\ and\ \bibinfo {author} {\bibfnamefont {R.~J.}\ \bibnamefont
  {Harrison}},\ }\href {https://doi.org/10.1063/1.462569} {\bibfield  {journal}
  {\bibinfo  {journal} {{J}.\ {C}hem.\ {P}hys.}\ }\textbf {\bibinfo {volume}
  {96}},\ \bibinfo {pages} {6796} (\bibinfo {year} {1992})}\BibitemShut
  {NoStop}%
\bibitem [{\citenamefont {Zaitsevskii}\ \emph {et~al.}(2020)\citenamefont
  {Zaitsevskii}, \citenamefont {Oleynichenko},\ and\ \citenamefont
  {Eliav}}]{Zaitsevskii:20}%
  \BibitemOpen
  \bibfield  {author} {\bibinfo {author} {\bibfnamefont {A.}~\bibnamefont
  {Zaitsevskii}}, \bibinfo {author} {\bibfnamefont {A.~V.}\ \bibnamefont
  {Oleynichenko}},\ and\ \bibinfo {author} {\bibfnamefont {E.}~\bibnamefont
  {Eliav}},\ }\href {https://doi.org/10.3390/sym12111845} {\bibfield  {journal}
  {\bibinfo  {journal} {{S}ymmetry}\ }\textbf {\bibinfo {volume} {12}},\
  \bibinfo {pages} {1845} (\bibinfo {year} {2020})}\BibitemShut {NoStop}%
\bibitem [{\citenamefont {Oleynichenko}\ \emph {et~al.}(2020)\citenamefont
  {Oleynichenko}, \citenamefont {Zaitsevskii},\ and\ \citenamefont
  {Eliav}}]{EXPT:20}%
  \BibitemOpen
  \bibfield  {author} {\bibinfo {author} {\bibfnamefont {A.~V.}\ \bibnamefont
  {Oleynichenko}}, \bibinfo {author} {\bibfnamefont {A.}~\bibnamefont
  {Zaitsevskii}},\ and\ \bibinfo {author} {\bibfnamefont {E.}~\bibnamefont
  {Eliav}},\ }in\ \href {https://doi.org/10.1007/978-3-030-64616-5{\_}33}
  {\emph {\bibinfo {booktitle} {Supercomputing}}},\ \bibinfo {series}
  {{C}ommun.\ {C}omput.\ {I}nf.\ {S}ci.}, Vol.\ \bibinfo {volume} {1331},\
  \bibinfo {editor} {edited by\ \bibinfo {editor} {\bibfnamefont
  {V.}~\bibnamefont {Voevodin}}\ and\ \bibinfo {editor} {\bibfnamefont
  {S.}~\bibnamefont {Sobolev}}}\ (\bibinfo  {publisher} {Springer International
  Publishing},\ \bibinfo {address} {Cham},\ \bibinfo {year} {2020})\ pp.\
  \bibinfo {pages} {375--386}\BibitemShut {NoStop}%
\bibitem [{DIR()}]{DIRAC:19}%
  \BibitemOpen
  \href@noop {} {}\bibinfo {note} {{DIRAC}, a relativistic ab initio electronic
  structure program, Release {DIRAC19} (2019), written by A.~S.~P.~Gomes,
  T.~Saue, L.~Visscher, H.~J.~{\relax Aa}.~Jensen, and R.~Bast, with
  contributions from I.~A.~Aucar, V.~Bakken, K.~G.~Dyall, S.~Dubillard,
  U.~Ekstr{\"o}m, E.~Eliav, T.~Enevoldsen, E.~Fa{\ss}hauer, T.~Fleig,
  O.~Fossgaard, L.~Halbert, E.~D.~Hedeg{\aa}rd, B.~Heimlich--Paris,
  T.~Helgaker, J.~Henriksson, M.~Ilia{\v{s}}, Ch.~R.~Jacob, S.~Knecht,
  S.~Komorovsk{\'y}, O.~Kullie, J.~K.~L{\ae}rdahl, C.~V.~Larsen, Y.~S.~Lee,
  H.~S.~Nataraj, M.~K.~Nayak, P.~Norman, G.~Olejniczak, J.~Olsen,
  J.~M.~H.~Olsen, Y.~C.~Park, J.~K.~Pedersen, M.~Pernpointner, R.~di~Remigio,
  K.~Ruud, P.~Sa{\l}ek, B.~Schimmelpfennig, B.~Senjean, A.~Shee, J.~Sikkema,
  A.~J.~Thorvaldsen, J.~Thyssen, J.~van~Stralen, M.~L.~Vidal, S.~Villaume,
  O.~Visser, T.~Winther, and S.~Yamamoto (available at
  {http://dx.doi.org/10.5281/zenodo.3572669}, see also
  {http://www.diracprogram.org})}\BibitemShut {NoStop}%
\bibitem [{\citenamefont {Saue}\ \emph {et~al.}(2020)\citenamefont {Saue},
  \citenamefont {Bast}, \citenamefont {Gomes}, \citenamefont {Jensen},
  \citenamefont {Visscher}, \citenamefont {Aucar}, \citenamefont {{Di
  Remigio}}, \citenamefont {Dyall}, \citenamefont {Eliav}, \citenamefont
  {Fasshauer}, \citenamefont {Fleig}, \citenamefont {Halbert}, \citenamefont
  {Hedeg\r{a}rd}, \citenamefont {{Helmich-Paris}}, \citenamefont {Ilia\v{s}},
  \citenamefont {Jacob}, \citenamefont {Knecht}, \citenamefont {Laerdahl},
  \citenamefont {Vidal}, \citenamefont {Nayak}, \citenamefont {Olejniczak},
  \citenamefont {Olsen}, \citenamefont {Pernpointner}, \citenamefont {Senjean},
  \citenamefont {Shee}, \citenamefont {Sunaga},\ and\ \citenamefont {{van
  Stralen}}}]{DIRAC:20}%
  \BibitemOpen
  \bibfield  {author} {\bibinfo {author} {\bibfnamefont {T.}~\bibnamefont
  {Saue}}, \bibinfo {author} {\bibfnamefont {R.}~\bibnamefont {Bast}}, \bibinfo
  {author} {\bibfnamefont {A.~S.~P.}\ \bibnamefont {Gomes}}, \bibinfo {author}
  {\bibfnamefont {H.~J.~A.}\ \bibnamefont {Jensen}}, \bibinfo {author}
  {\bibfnamefont {L.}~\bibnamefont {Visscher}}, \bibinfo {author}
  {\bibfnamefont {I.~A.}\ \bibnamefont {Aucar}}, \bibinfo {author}
  {\bibfnamefont {R.}~\bibnamefont {{Di Remigio}}}, \bibinfo {author}
  {\bibfnamefont {K.~G.}\ \bibnamefont {Dyall}}, \bibinfo {author}
  {\bibfnamefont {E.}~\bibnamefont {Eliav}}, \bibinfo {author} {\bibfnamefont
  {E.}~\bibnamefont {Fasshauer}}, \bibinfo {author} {\bibfnamefont
  {T.}~\bibnamefont {Fleig}}, \bibinfo {author} {\bibfnamefont
  {L.}~\bibnamefont {Halbert}}, \bibinfo {author} {\bibfnamefont {E.~D.}\
  \bibnamefont {Hedeg\r{a}rd}}, \bibinfo {author} {\bibfnamefont
  {B.}~\bibnamefont {{Helmich-Paris}}}, \bibinfo {author} {\bibfnamefont
  {M.}~\bibnamefont {Ilia\v{s}}}, \bibinfo {author} {\bibfnamefont {C.~R.}\
  \bibnamefont {Jacob}}, \bibinfo {author} {\bibfnamefont {S.}~\bibnamefont
  {Knecht}}, \bibinfo {author} {\bibfnamefont {J.~K.}\ \bibnamefont
  {Laerdahl}}, \bibinfo {author} {\bibfnamefont {M.~L.}\ \bibnamefont {Vidal}},
  \bibinfo {author} {\bibfnamefont {M.~K.}\ \bibnamefont {Nayak}}, \bibinfo
  {author} {\bibfnamefont {M.}~\bibnamefont {Olejniczak}}, \bibinfo {author}
  {\bibfnamefont {J.~M.~H.}\ \bibnamefont {Olsen}}, \bibinfo {author}
  {\bibfnamefont {M.}~\bibnamefont {Pernpointner}}, \bibinfo {author}
  {\bibfnamefont {B.}~\bibnamefont {Senjean}}, \bibinfo {author} {\bibfnamefont
  {A.}~\bibnamefont {Shee}}, \bibinfo {author} {\bibfnamefont {A.}~\bibnamefont
  {Sunaga}},\ and\ \bibinfo {author} {\bibfnamefont {J.~N.~P.}\ \bibnamefont
  {{van Stralen}}},\ }\href {https://doi.org/10.1063/5.0004844} {\bibfield
  {journal} {\bibinfo  {journal} {{J}.\ {C}hem.\ {P}hys.}\ }\textbf {\bibinfo
  {volume} {152}},\ \bibinfo {pages} {204104} (\bibinfo {year}
  {2020})}\BibitemShut {NoStop}%
\bibitem [{\citenamefont {Skripnikov}(2021)}]{Skripnikov:2021}%
  \BibitemOpen
  \bibfield  {author} {\bibinfo {author} {\bibfnamefont {L.~V.}\ \bibnamefont
  {Skripnikov}},\ }\href {https://doi.org/10.1063/5.0053659} {\bibfield
  {journal} {\bibinfo  {journal} {The Journal of Chemical Physics}\ }\textbf
  {\bibinfo {volume} {154}},\ \bibinfo {pages} {201101} (\bibinfo {year}
  {2021})}\BibitemShut {NoStop}%
\bibitem [{\citenamefont {Pazyuk}\ \emph {et~al.}(2015)\citenamefont {Pazyuk},
  \citenamefont {Zaitsevskii}, \citenamefont {Stolyarov}, \citenamefont
  {Tamanis},\ and\ \citenamefont {Ferber}}]{Pazyuk:15}%
  \BibitemOpen
  \bibfield  {author} {\bibinfo {author} {\bibfnamefont {E.~A.}\ \bibnamefont
  {Pazyuk}}, \bibinfo {author} {\bibfnamefont {A.~V.}\ \bibnamefont
  {Zaitsevskii}}, \bibinfo {author} {\bibfnamefont {A.~V.}\ \bibnamefont
  {Stolyarov}}, \bibinfo {author} {\bibfnamefont {M.}~\bibnamefont {Tamanis}},\
  and\ \bibinfo {author} {\bibfnamefont {R.}~\bibnamefont {Ferber}},\ }\href
  {https://doi.org/10.1070/RCR4534} {\bibfield  {journal} {\bibinfo  {journal}
  {{R}uss.\ {C}hem.\ {R}ev.}\ }\textbf {\bibinfo {volume} {84}},\ \bibinfo
  {pages} {1001} (\bibinfo {year} {2015})}\BibitemShut {NoStop}%
\bibitem [{\citenamefont {Zaitsevskii}\ \emph {et~al.}(2017)\citenamefont
  {Zaitsevskii}, \citenamefont {Mosyagin}, \citenamefont {Stolyarov},\ and\
  \citenamefont {Eliav}}]{Zaitsevskii:17}%
  \BibitemOpen
  \bibfield  {author} {\bibinfo {author} {\bibfnamefont {A.}~\bibnamefont
  {Zaitsevskii}}, \bibinfo {author} {\bibfnamefont {N.~S.}\ \bibnamefont
  {Mosyagin}}, \bibinfo {author} {\bibfnamefont {A.~V.}\ \bibnamefont
  {Stolyarov}},\ and\ \bibinfo {author} {\bibfnamefont {E.}~\bibnamefont
  {Eliav}},\ }\href {https://doi.org/10.1103/PhysRevA.96.022516} {\bibfield
  {journal} {\bibinfo  {journal} {{P}hys.\ {R}ev.\ {A}}\ }\textbf {\bibinfo
  {volume} {96}},\ \bibinfo {pages} {022516} (\bibinfo {year}
  {2017})}\BibitemShut {NoStop}%
\end{thebibliography}%

\end{document}